\documentclass[a4paper,superscriptaddress,notitlepage,12pt,tightenlines]{revtex4-2}
\usepackage{graphicx}
\usepackage{amsmath,amssymb,setspace}
\usepackage{newtxtext}
\usepackage{newtxmath}
\usepackage{ulem}
\usepackage{verbatim}
\usepackage[x11names]{xcolor}
\usepackage[unicode,colorlinks,citecolor=Blue3,linkcolor=Blue3,bookmarks=true]{hyperref}

\newcommand{\eg}{{\it e.g.\,}}

\newcommand{\fulld}[2]{\dfrac{d#1}{d#2}}
\newcommand{\intinfty}{\displaystyle\int_{-\infty}^{\infty}\!}
\newcommand{\mean}[1]{\langle{#1}\rangle}
\newcommand{\svector}[2]{\begin{pmatrix}#1 \\ #2 \end{pmatrix}}
\newcommand{\smatrix}[4]{\begin{pmatrix}#1 & #2 \\ #3 & #4\end{pmatrix}}

\begin{document}

\title{Sensitivity limits of non-stationary quantum sensors}

\date{\today}

\author{Farid Ya.\ Khalili}
\email{farit.khalili@gmail.com}
\affiliation{Russian Quantum Center, Skolkovo IC, Bolshoy Bulvar 30, bld.\ 1, Moscow, 121205, Russia}

\begin{abstract}

The concept of the dissipative quantum limit (DQL) was first put forward in 1980s and was analyzed in detail much later in Ref. [Phys. Rev. A 103, 043721 (2021)] for the particular case of stationary (invariant with respect to a shift of time) systems. Here we extend that analysis to the general non-stationary case.

\end{abstract}

\maketitle

\section{Introduction}\label{sec:intro}

The sensitivity of the best modern optomechanical force sensors is limited by quantum effects. For example, the laser interferometric gravitational-wave (GW) detectors \cite{CQG_32_7_074001_2015, Acernese_CQG_32_024001_2015, Tse_PRL_123_231107_2019_short, Acernese_PRL_123_231108_2019_short} have reached a sensitivity which is close to the Standard Quantum Limit (SQL), which correponds to the balance of the measurement imprecision and perturbation of the probe mirrors by the meter's back action noise originating from the Heisenberg uncertainty relation \cite{67a1eBr, 74a1eBrVo, Caves_RMP_52_341_1980, 92BookBrKh}. Being expressed as the spectral density of the effective force noise, it has the following form, see \cite{12a1DaKh}:
\begin{equation}\label{SQL}
  S_{\rm SQL}(\Omega) = \hbar|\chi^{-1}(\Omega)| \,,
\end{equation}
where $\chi^{-1}$ is the response function of the probe object (for example, for a free mass, $\chi^{-1}(\Omega)=-m\Omega^2$).

The SQL is not a truly fundamental limit and can be evaded, in particular, using a meter with properly anti-correlated measurement noise and back action noise \cite{Unruh1982, 02a1KiLeMaThVy} (see also the review articles~\cite{12a1DaKh, 19a1DaKhMi}). Recently this approach was implemented in the LIGO GW detectors \cite{Ganapathy_PRX_13_041021_2023, Jia_Science_385_1318_2024}. Overcoming the SQL using similar technologies was also demonstrated in smaller scale (table-top) optomechanical setups, see \eg Refs.\,\cite{Kampel_PRX_7_021008_2017, Moeller_Nature_547_191_2017, Mason_NPhys_15_745_2019}.

Another approach to overcoming the SQL is the use of non-stationary systems, that is the ones whose dynamic or noise properties (or both) explicitly depend on time. Actually, the first proposed schemes of overcoming the SQL \cite{78a1eBrKhVo, Thorne1978, Caves_RMP_52_341_1980} belong to this category. We would like to mention also the relatively recent works \cite{Woolley_PRA_87_063846_2013, Buchmann_PRL_117_030801_2016, Ockeloen-Korppi_PRL_117_140401_2016, 22a1ZePoKh}

Two quantum sensitivity limitations which are more fundamental than the SQL are known. The first one arises because in the real-world meters, the measurements strength (in the
of optical interferometers, the circulating optical power) is limited. This constrain is known as the energetic quantum limit \cite{00p1BrGoKhTh} and in more general context corresponds to the quantum Cramér-Rao bound (QCRB) \cite{HelstromBook, Miao_PRL_119_050801_2017, Tsang_PRL_106_090401_2011}.

The second limit originates from the dissipative dynamics of the probe object, that is from the imaginary part $\Im\chi^{-1}$ of its response function. It arises because the dissipative dynamics does not allow for full suppression of the quantum noise using the anti-correlation mechanism of Ref.\,\cite{Unruh1982}. In the case of stationary (invariant with respect to a shift of time) systems, this gives the following limitation, see Ref.\,\cite{87a1eKh}:
\begin{equation}\label{DQL}
  S_{\rm DQL}(\Omega) = \hbar|\Im\chi^{-1}(\Omega)| \,.
\end{equation}
It was explored in detail in Ref.\,\cite{20a1KhZe}, where the term ``dissipative quantum limit'' (DQL) was proposed for it. Note that the spectral density \eqref{DQL} should not be confused with the thermal noise spectral density which, for the paricular case of a zero temperature, has the same form. They originate from different noise sources, namely the meter quantum noises and the probe object thermal noise.

In the same work \cite{20a1KhZe}, the applicability of the DQL to non-stationary systems, that is the ones whose dynamic or noise properties (or both) explicitly depend on time was also discussed briefly. It was concluded in the general non-stationary case, it should be equal to zero.

The goal of this is to provide the rigorous analysis of this problem. The paper organized as follows. In the next introductory section, we remind the reader the basic principles of linear measurement theory and introduce the main notations used throughout the paper. In Sec.\,\ref{sec:DQL}, we explore the applicability of the DQL to the non-stationary case. Finally, in Sec.\,\ref{sec:conclusion} we resume the results of this work.

\section{Linear probe system}\label{sec:linear}

The very small values of the signals typical for the quantum sensors allow to consider them as linear devices. This feature radically simplify their theoretical analysis. In addition, in the stationary case, the Fourier representation can be used, and actually is used almost exclusively, further simplifying the analysis of the linear stationary sensors.
In particular, the SQL \eqref{SQL} and DQL \eqref{DQL} were obtained using this approach.

\begin{figure}[t]
\includegraphics[scale=1]{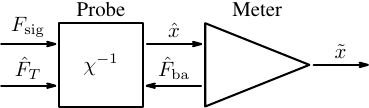}
\caption{Generic linear force sensor consisting of probe and meter subsystems. The probe is subjected to a classical signal force $F_{\text{sig}}$, a thermal force $\hat{F}_T$ due to dissipation, and a back-action force $\hat{F}_{\text{ba}}$ from the meter. $\chi^{-1}$ is the probe responce function, $\tilde{x}$ is the meter output normalized to the probe position.}
\label{fig:linear}
\end{figure}

Let us consider the general scheme of the linear probe system, shown in Fig.\,\ref{fig:linear}. Here the signal force $F_{\rm sig}$ which has to be detected acts on the linear probe, described by the susceptibility function $\chi(t,t')$. The coordinate $\hat{x}$ of the probe is measured by the linear meter, which output signal $\tilde{y}(t)$ is propotional to the sum of $\hat{x}$ and the measurement see noise $\hat{x}_{{\rm fl}}$:
\begin{equation}
  \tilde{y}(t) = \intinfty G(t,t')\bigl(\hat{x}(t') + \hat{x}_{\rm fl}(t')\bigr)dt' \,,
\end{equation}
where $G(t,t')$ is the causal transfer function of the meter. Evidently, it can be removed by the data processing of $\tilde{y}$ (which can be non-causal), giving the signal reduced to the meter input:
\begin{equation}\label{meter_out}
  \tilde{x}(t) = \hat{x}(t) + \hat{x}_{\rm fl}(t) \,.
\end{equation}

At the same time, the meter applies the back action force $\hat{F}_{{\rm ba}}$ to the probe, consisting of the dynamic part, proportional to $\hat{x}$, and the back action noise $\hat{F}_{{\rm fl}}$:
\begin{equation}\label{F_ba}
  \hat{F}_{{\rm ba}}(t) = \hat{F}_{{\rm fl}}(t) - \intinfty K(t,t')\hat{x}(t')\,dt' \,,
\end{equation}
where $K$ is the dynamic rigidity introduced by the meter. In the optical interferometers case, this effect is known as the optical spring~\cite{99a1BrKh, Buonanno2002, 12a1DaKh}.

The corresponding equation of motion of the probe is the following:
\begin{equation}\label{probe}
  \intinfty\chi^{-1}_K(t,t')\hat{x}(t')\,dt'
  = F_{{\rm sig}}(t) + \hat{F}_{\rm fl}(t) + \hat{F}_T(t) \,,
\end{equation}
where $\hat{F}_T$ is the probe thermal noise,
\begin{equation}
  \chi^{-1}_K(t,t') = \chi^{-1}(t,t') + K(t,t')
\end{equation}
is the probe response function modified by the rigidity $K$, and for any kernel $\kappa(t,t')$ its inverse $\kappa^{-1}$ is defined as follow:
\begin{equation}\label{inv_kernel}
  \intinfty \kappa^{-1}(t,t'')\kappa(t'',t')dt'' = \delta(t-t') \,.
\end{equation}

Combining Eqs.\,\eqref{meter_out} and \eqref{probe}, we obtain that:
\begin{equation}
  \tilde{x}(t) = \hat{x}_{\rm fl}(t)  + \intinfty\chi_K(t',t'')\bigl(
      F_{\rm sig}(t'') + \hat{F}_{\rm fl}(t'') + \hat{F}_T(t'')
    \bigr)dt' \,.
\end{equation}
We assume then that $\tilde{x}$ is further data-processed to obtain the estimate of the signal force $\tilde{F}$:
\begin{equation}\label{tilde_F}
  \tilde{F}(t) = \intinfty D_K(t,t')\tilde{x}(t')\,dt'
  = F_{\rm sig}(t) + \hat{F}_{\rm sum}(t) + \hat{F}_T(t) \,,
\end{equation}
where
\begin{equation}\label{F_sum}
  \hat{F}_{\rm sum}(t)
  = \intinfty\chi_K^{-1}(t,t')\hat{x}_{\rm fl}(t')\,dt' + \hat{F}_{\rm fl}(t)
\end{equation}
is the sum noise of the meter.

It is known that response functions of the linear quantum systems and their noises commutators depend on each other \cite{Kubo1956}. In the particular case of the linear meter considered here, this gives the following relations, see Ref.\,\cite{82a2eKhVo}:
\begin{subequations}\label{commutators}
  \begin{gather}
    [\hat{x}_{\rm fl}(t),\hat{x}_{\rm fl}(t')] = 0 \,, \\
    [\hat{F}_{\rm fl}(t),\hat{F}_{\rm fl}(t')] = i\hbar\bigl(K(t,t') - K(t',t)\bigr) \,, \\
    [\hat{x}_{\rm fl}(t),\hat{F}_{\rm fl}(t')] = -i\hbar\delta(t-t') \,,
  \end{gather}
\end{subequations}
\begin{equation}\label{comm_T}
  [\hat{F}_T(t),\hat{F}_T(t')] = 2i\hbar\chi_a^{-1}(t,t') \,,
\end{equation}
where
\begin{equation}
  \chi_a^{-1}(t,t') = \tfrac12\bigl(\chi^{-1}(t,t') - \chi^{-1}(t',t)\bigr)
\end{equation}
is the antisymmetric part of the $\chi^{-1}$ responsible for the dissipation.

In the general non-stationary case, we quantify the noises by the symmetrized correlation functions defined as follows:
\begin{subequations}\label{comp_noises}
  \begin{gather}
    B_{xx}(t,t') = \mean{\hat{x}_{\rm fl}(t)\circ\hat{x}_{\rm fl}(t')} \,, \\
    B_{FF}(t,t') = \mean{\hat{F}_{\rm fl}(t)\circ\hat{F}_{\rm fl}(t')} \,, \\
    B_{xF}(t,t') = B_{Fx}(t',t) = \mean{\hat{x}_{\rm fl}(t)\circ\hat{F}_{\rm fl}(t')} \,,
  \end{gather}
\end{subequations}
\begin{equation}
  B_{TT}(t,t') = \mean{\hat{F}_T(t)\circ\hat{F}_T(t')} \,,
\end{equation}
where $``\circ''$ mean symmetric product: for any $\hat{Q}$, $\hat{P}$,
\begin{equation}
  \hat{Q}\circ\hat{P} = \tfrac12(\hat{Q}\hat{P} + \hat{P}\hat{Q}) .
\end{equation}
It can be shown using the straightforward calculation that the correlation function of the sum noise \eqref{F_sum} is equal to
\begin{multline}\label{B_sum}
  B_{\rm sum}(t,t')
  = \intinfty\chi_K^{-1}(t,t_1)\chi_K^{-1}(t',t_1')B_{xx}(t_1,t_1')\,dt_1dt_1'
    + \intinfty\chi_K^{-1}(t,t_1)B_{xF}(t_1,t')\,dt_1 \\
    + \intinfty\chi_K^{-1}(t',t_1)B_{xF}(t_1,t)\,dt_1 + B_{FF}(t,t') \,.
\end{multline}

It was shown also in Ref\,\cite{92BookBrKh} using the commutators \eqref{commutators} that the following uncertainty relation for the correlation functions \eqref{comp_noises} is valid for any quantum linear meter:
\begin{multline}\label{BxBF}
  \intinfty\bigl(Q_F^*(t)\ Q_x^*(t)\bigr)
    \smatrix{B_{FF}(t,t') + \dfrac{i\hbar}{2}\bigl(K(t,t') - K(t',t))\bigr)\ }
      {B_{Fx}(t,t') + \dfrac{i\hbar}{2}\delta(t-t')}
      {B_{xF}(t,t') - \dfrac{i\hbar}{2}\delta(t-t')}
      {B_{xx}(t,t')} \\ \times
    \svector{Q_F(t')}{Q_x(t')}dtdt' \ge 0 \,,
\end{multline}
where $Q_F(t)$, $Q_x(t)$ are two arbitrary complex-valued functions.

It the rest of this section, we consider the stationary system. This means two conditions. First, its dynamic properties are invariant under the shift of time:
\begin{equation}\label{stat_chi}
  \chi^{-1}(t+\tau,t) = \chi^{-1}(\tau,0)
  = \intinfty\chi^{-1}(\Omega)e^{-i\Omega\tau}\,\frac{d\Omega}{2\pi}\,,
\end{equation}
and similarly for $K$ and $\chi_K^{-1}$. Second, the same is true for the noise correlaton functions:
\begin{equation}
  B_{\alpha\beta}(t+\tau,t) = B_{\alpha\beta}(\tau,0)
  = \intinfty S_{\alpha\beta}(\Omega)e^{-i\Omega\tau}\frac{d\Omega}{2\pi} \,,
\end{equation}
where $\alpha,\beta=x,F,T$ and $S_{\alpha\beta}$ are the respective spectral densities.

In this case, it follows from Eq.\,\eqref{B_sum} that the spectral density of the sum noise \eqref{F_sum} is equal to
\begin{equation}\label{S_sum}
  S_{\rm sum}(\Omega) = |\chi^{-1}_K(\Omega)|^2S_{xx}(\Omega)
    + 2\Re\bigl(\chi^{-1}_K(\Omega)S_{xF}(\Omega)\bigr) + S_{FF}(\Omega) \,,
\end{equation}
and the inequalality \eqref{BxBF} reduces to the following one, see Ref.\,\cite{82a2eKhVo}:
\begin{equation}\label{SxSF}
  S_{xx}(\Omega)S_{FF}(\Omega) - |S_{xF}(\Omega)|^2
    \ge \hbar\bigl|\Im\bigl(K^*(\Omega)S_{xx}(\Omega) + S_{xF}(\Omega)\bigr)\bigr|
    + \frac{\hbar^2}{4} \,.
\end{equation}
Rigorous optimization of the sum noise spectral density \eqref{S_sum} under the condition \eqref{SxSF} gives the DQL \eqref{DQL}.

\section{Non-stationary systems}\label{sec:DQL}

\paragraph{The sum noise.}

For the non-stationary meters, a simple closed form of the uncertainty relation for $\hat{x}_{\rm fl}$ and $\hat{F}_{\rm fl}$, similar to Eq.\,\eqref{SxSF}, is unknown.  Therefore, here we explore the sensitivity of the non-stationary meters starting directly from the commutators \eqref{commutators}.

It was conjectured in Ref.\,\cite{20a1KhZe} that because the meter output signal $\tilde{F}$ (see Eq.\,\eqref{tilde_F}) is a classical observable, its autommutator has to be equal to zero. Using Eqs.\,\eqref{commutators}, it can be shown explicitly that this in indeed the case. Really, the autocommutator of the sum noise \eqref{F_sum} is equal to
\begin{multline}\label{F_sum_comm}
  [\hat{F}_{\rm sum}(t),\hat{F}_{\rm sum}(t')] \\
  = \intinfty \chi_K^{-1}(t,t'')
      [\hat{x}_{\rm fl}(t''),\hat{F}_{\rm fl}(t')]\,dt'' 
    + \intinfty \chi_K^{-1}(t',t'')
        [\hat{F}_{\rm fl}(t),\hat{x}_{\rm fl}(t'')]\,dt''
    + [\hat{F}_{\rm fl}(t), \hat{F}_{\rm fl}(t')] \\
  = i\hbar\bigl(-\chi_K^{-1}(t,t') + \chi_K^{-1}(t',t) + K(t,t') - K(t',t)\bigr)
  = -2i\hbar\chi_a^{-1}(t,t') \,.
\end{multline}
It follows from Eqs.\,\eqref{comm_T} and \eqref{F_sum_comm} that
\begin{equation}\label{comm_tilde_F}
  [\tilde{F}(t),\tilde{F}(t')]
  = [\hat{F}_{\rm sum}(t),\hat{F}_{\rm sum}(t')] + [\hat{F}_T(t),\hat{F}_T(t')] = 0 \,.
\end{equation}
As it was mentioned in Ref.\,\cite{20a1KhZe}, this result also means that in the stationary case, spectral density of $\hat{F}_{\rm sum}$ can not be smaller than \eqref{DQL}, providing thus a simple ``shortcut'' proof of the DQL.

\paragraph{Detection of a given force.}

Consider now the problem of detection of a signal force with a priory known shape $F_{\rm sig}(t)$. The sensitivity in this case is defined by the signal-to-noise ratio
\begin{equation}
  {\rm SNR}
  = \frac{1}{\mean{\mathcal{\hat{F}}_{\rm sum}^2} + \mean{\mathcal{\hat{F}}_T^2}}
      \biggl(\intinfty\Phi(t)F_{\rm sig}(t)\,dt\biggr)^2 ,
\end{equation}
where $\Phi(t)$ is a filter function,
\begin{subequations}
  \begin{gather}
    \mathcal{\hat{F}}_{\rm sum} = \intinfty\Phi(t)\hat{F}_{\rm sum}(t)\,dt \,, \\
    \mathcal{\hat{F}}_T = \intinfty\Phi(t)\hat{F}_T(t)\,dt
  \end{gather}
\end{subequations}
are the measurement errors introduced by the meter sum noise and the probe thermal noise, and
 \begin{equation}
  \mean{\mathcal{\hat{F}}_{\rm sum}^2} = \intinfty\Phi(t)B_{\rm sum}(t,t') \Phi(t')dtdt' \,.
\end{equation}

For any $\Phi(t)$, in the absence of the stationarity constraints, the meter can be prepared in a quantum state with a given value of $\mathcal{\hat{F}}_{\rm sum}$, which, without limiting the generality, can be assumed equal to zero. This means that the DQL does not affect the sensitivity in this case.

In order to further confirm this result, and taking into account that one example is sufficient for this, we consider a simple particular case of memoryless  meter. Namely, suppose that the correlation functions \eqref{comp_noises} have the form of $\delta$-functions with time-dependent prefactors:
\begin{equation}\label{B_delta}
  B_{xx}(t,t') = S_{xx}(t)\delta(t-t') \,,\quad
  B_{FF}(t,t') = S_{FF}(t)\delta(t-t') \,,\quad
  B_{xF}(t,t') = S_{xF}(t)\delta(t-t') \,.
\end{equation}
We assume also for simplicity that the factor $K=0$.

It follows from Eqs.\,\eqref{B_sum} that in this case,
\begin{equation}\label{deltaF}
  \mean{\mathcal{\hat{F}}_{\rm sum}^2}
  = \intinfty\bigl(\Psi^2(t)S_{xx}(t) + 2\Psi(t)\Phi(t)S_{xF}(t) + \Phi^2(t)S_{FF}(t)\bigr)
      dt \,,
\end{equation}
where
\begin{equation}
  \Psi(t) = \intinfty\Phi(t')\chi^{-1}(t',t)\,dt' \,.
\end{equation}
The uncertainty relation \eqref{BxBF} in the case of \eqref{B_delta} takes the following simple form:
\begin{equation}\label{SxSFnonstat}
  S_{xx}(t)S_{FF}(t) - S_{xF}^2(t) \ge \frac{\hbar^2}{4} \,.
\end{equation}

Let us minimize now Eq.\,\eqref{deltaF} under the constraint \eqref{SxSFnonstat}, assuming the exact equality in the last equation. We assume also that the function $S_{FF}$ is fixed (note that it is proportional to the measurements strength, for example to the optical power in the interferometer). It is easy to show that in this case, the minimum of \eqref{deltaF}, for any given filter function $\Phi$, is achieved at
\begin{equation}
  S_{xF}(t) = -\frac{\Phi(t)}{\Psi(t)}S_{FF}(t)
\end{equation}
and is equal to
\begin{equation}
  \mean{\mathcal{\hat{F}}_{\rm sum}^2}
  = \frac{\hbar^2}{4}\intinfty\frac{\Psi^2(t)}{S_{FF}(t)}\,dt
\end{equation}
This result has the form typical for the QCRB, compare \eg with Eq.\,(8) of \cite{20a1KhZe}. At the same time, it is not affected by the DQL and in the (hypothetical) case of $S_{FF}\to\infty$, the value of $\mean{\mathcal{\hat{F}}_{\rm sum}^2}$ can be reduced to zero.

\paragraph{Measurement of two quadratures of the signal force.}

The situation is different in the case of several possible shapes of the signal force. In this case, a set of the filter functions $\Phi_j$, where $j=1,2,\dots$, should be used \cite{HelstromBook}, and the cancellation of the meter noise for all values of $j$ at once is possible only if the integrals
\begin{equation}
  \mathcal{\hat{F}}_{\rm sum}^{(j)} = \intinfty\Phi_j(t)\hat{F}_{\rm sum}(t)\,dt
\end{equation}
commute with each other. It follows from Eq.\,\eqref{F_sum_comm} that the corresponding commutators are equal to
\begin{equation}\label{comm_jk}
  [\mathcal{\hat{F}}_{\rm sum}^{(j)},\mathcal{\hat{F}}_{\rm sum}^{(k)}]
  = -2i\hbar\intinfty\Phi_j(t)\chi_a^{-1}(t,t')\Phi_k(t')\,dtdt' \,.
\end{equation}
In the general non-commuting case, they gives the following uncertainty relations:
\begin{equation}\label{deltas_jk}
  \mean{(\mathcal{\hat{F}}_{\rm sum}^{(j)})^2}\mean{(\mathcal{\hat{F}}_{\rm sum}^{(k)})^2}
  \ge \hbar^2\biggl(\intinfty\Phi_j(t)\chi_a^{-1}(t,t')\Phi_k(t')\,dtdt'\biggr)^2 \,.
\end{equation}

As a practical example, consider measurement of quadrature amplitudes $F_c$, $F_s$ of a narrow-band force of the following form:
\begin{equation}\label{nb_force}
  F_{\rm sig}(t) = F_c(t)\cos\Omega_0t + F_s(t)\sin\Omega_0t \,,
\end{equation}
where the functions $F_{c,s}$ are slow-varying ones, in comparison with the frequency $\Omega_0$. Note that non-stationary schemes allowing to measure one of the two quadrature (but not both) were discussed, in, particular, in Refs.\,\cite{78a1eBrKhVo, Thorne1978, Caves_RMP_52_341_1980, Buchmann_PRL_117_030801_2016, Ockeloen-Korppi_PRL_117_140401_2016}.

We suppose that the dissipation in the probe object is stationary. In the real-world scenarios,  typically, this is the case. This assumption allows to present the antisymmetric part of the response function in the following form:
\begin{equation}
  \chi_a^{-1}(t-t') = \intinfty\chi_a^{-1}(\Omega)e^{-i\Omega(t-t')}\frac{d\Omega}{2\pi} \,.
\end{equation}
In the narrow-band case of \eqref{nb_force}, the function $\chi_a^{-1}(\Omega)$ can be approximated as follows:
\begin{equation}
  \chi_a^{-1}(\Omega) = -i\Omega H
\end{equation}
where $H$ is the friction coefficient, giving that
\begin{equation}
  \chi_a^{-1}(t-t') = H\fulld{\delta(t-t')}{t} \,,
\end{equation}
In this case, replacing in Eqs.\,\eqref{comm_jk}, \eqref{deltas_jk} the supercripts $(j),(k)$ by $c,s$, we obtain:
\begin{equation}\label{comm_cs}
  [\mathcal{\hat{F}}_{\rm sum}^c, \mathcal{\hat{F}}_{\rm sum}^s]
  = -2i\hbar H\intinfty\fulld{\Phi_c(t)}{t}\Phi_s(t)\,dt \,.
\end{equation}

Evidently, in order to measure the force quadratures, the following filter functions should be used:
\begin{equation}
  \Phi_c(t) = \Phi_{c0}(t)\cos\Omega_0t\,, \quad \Phi_s(t) = \Phi_{s0}\sin\Omega_0t \,,
\end{equation}
where $\Phi_{c0, s0}$ are slow-varying in comparison with the frequency $\Omega_0$ amplitudes. Substitution of these functions into Eq.\,\eqref{comm_jk} gives that
\begin{equation}\label{comm_cs}
  [\mathcal{\hat{F}}_{\rm sum}^c, \mathcal{\hat{F}}_{\rm sum}^s]
  \approx 2i\hbar\Omega_0H\intinfty\Phi_{c0}(t)\Phi_{s0}(t)\sin^2\Omega_0t\,dt
  \approx i\hbar\Omega_0H\intinfty\Phi_{c0}(t)\Phi_{s0}(t)\,dt \,,
\end{equation}
As a result, we obtain following the DQL-like limit:
\begin{equation}\label{deltas_cs}
  \mean{(\mathcal{\hat{F}}_{\rm sum}^c)^2}\mean{(\mathcal{\hat{F}}_{\rm sum}^s)^2}
  \ge\frac{\hbar^2|\chi_a^{-1}(\Omega_0)|^2}{4}
    \biggl(\intinfty\Phi_{c0}(t)\Phi_{s0}(t)\,dt\biggr)^2 \,.
\end{equation}

\section{Conclusion}\label{sec:conclusion}

The main results of this paper can be formulated as follows. We calculated the autocommutator of the sum quantum noise of a non-stationary linear force sensor, see Eq.\,\eqref{F_sum_comm}, and showed explicitly that the autocommutator of the output signal of the meter vanishes, as it was predicted in Ref.\,\cite{20a1KhZe}.

We showed also that the sensitivity to the signal force with a priory known shape is not limited by the DQL. At the same time, in the case of several possible shapes of the signal force, the sensitivity is limited by the set of DQL-like uncertinty relations depending on the antysymmetric part $\chi_a^{-1}$, see Eq.\,\eqref{deltas_jk}

As a practical example, we considered the simultaneous measurement of two quadratures amplitudes of a narrow band force. We derived the specific for this particular case form of the above mentioned DQL-like uncertainty relation, see Eq.\,\eqref{deltas_cs}.

The following interesting conclusion follows from the uncertainty relations \eqref{deltas_jk}, \eqref{deltas_cs}. They does not depend on the full response function $\chi^{-1}$ of the probe object, which appears in the SQL. Therefore, in principle, several parameters of the signal force (for example, both quadratures) can be measured simultaneously with the precision not limited by the SQL, provided that the quantum noise of the meter is optimized properly.

\acknowledgements

This work was supported by the Foundation for the Advancement of Theoretical Physics and Mathematics ``BASIS'', Grant 23-1-1-39-1.


%

\end{document}